\documentclass[conference]{IEEEtran}
\IEEEoverridecommandlockouts
\usepackage{cite}
\usepackage{amssymb,amsfonts}
\usepackage{algorithmic}
\usepackage{graphicx}
\usepackage{tikz}
\usetikzlibrary{positioning}
\usepackage{textcomp}
\usepackage{xcolor}
\usepackage{bm}
\usepackage[]{amsmath}
\usepackage[caption=false]{subfig}
\def\BibTeX{{\rm B\kern-.05em{\sc i\kern-.025em b}\kern-.08em
    T\kern-.1667em\lower.7ex\hbox{E}\kern-.125emX}}
\begin{document}

\title{Evolution of Collective Decision-Making Mechanisms for Collective Perception
\thanks{This work was partially supported by the German Federal Ministry of Education and Research (BMBF, SCADS22B) and the Saxon State Ministry for Science, Culture and Tourism (SMWK) by funding the competence center for Big Data and AI ``ScaDS.AI Dresden/Leipzig''. \\[1em]
\copyright2023 IEEE. Personal use of this material is permitted. Permission from IEEE must be obtained for all other uses, in any current or future media, including reprinting/republishing this material for advertising or promotional purposes, creating new collective works, for resale or redistribution to servers or lists, or reuse of any copyrighted component of this work in other works.}}

\author{\IEEEauthorblockN{Tanja Katharina Kaiser}
\IEEEauthorblockA{\textit{TU Dresden} \& \\
\textit{ScaDS.AI Dresden/Leipzig}\\
Dresden, Germany \\
0000-0002-1700-5508}
\and
\IEEEauthorblockN{Tristan Potten}
\IEEEauthorblockA{\textit{Institute of Computer Engineering} \\
\textit{University of Lübeck}\\
Lübeck, Germany \\
0000-0003-4873-0398}
\and 
\IEEEauthorblockN{Heiko Hamann}
\IEEEauthorblockA{\textit{Dept. of Computer \& Information Science} \\
\textit{University of Konstanz}\\
Konstanz, Germany \\
0000-0002-2458-8289}
}

\IEEEoverridecommandlockouts

\maketitle

\IEEEpubidadjcol

\begin{abstract}
Autonomous robot swarms must be able to make fast and accurate collective decisions, but speed and accuracy are known to be conflicting goals.
While collective decision-making is widely studied in swarm robotics research, only few works on using methods of evolutionary computation to generate collective decision-making mechanisms exist. 
These works use task-specific fitness functions rewarding the accomplishment of the respective collective decision-making task. 
But task-independent rewards, such as for prediction error minimization, may promote the emergence of diverse and innovative solutions. 
We evolve collective decision-making mechanisms using a task-specific fitness function rewarding correct robot opinions, a task-independent reward for prediction accuracy, and a hybrid fitness function combining the two previous. 
In our simulations, we use the collective perception scenario, that is, robots must collectively determine which of two environmental features is more frequent. 
We show that evolution successfully optimizes fitness in all three scenarios, but that only the task-specific fitness function and the hybrid fitness function lead to the emergence of collective decision-making behaviors. 
In benchmark experiments, we show the competitiveness of the evolved decision-making mechanisms to the voter model and the majority rule and analyze the scalability of the decision-making mechanisms with problem difficulty. 
\end{abstract}


\section{Introduction}

Robot swarms~\cite{Hamann2018} are decentralized multi-robot systems that rely on local communication and perception. 
The global swarm behavior emerges from local interactions between robots and between robots and their environment.
Collective decision-making is the capability of a robot swarm to make a choice among multiple options as a collective and essential to creating fully autonomous swarm robotic systems.
A~frequently used problem scenario is collective perception~\cite{Valentini2016} where swarms must determine which of two environmental features is more frequent. 
Over the years, a variety of collective decision-making mechanisms have been proposed, the most well-known ones are the voter model and the majority rule~\cite{Hamann2018}. 
To make a collective decision, swarm members adopt the opinion of a random neighbor in the voter model. 
By contrast, swarm members adopt the majority opinion of their neighbors when using the majority rule.
These two decision-making mechanisms illustrate the so-called speed versus accuracy trade-off, that is, faster decision-making leads to a loss of decision accuracy and vice versa~\cite{Valentini2015}. 
On average, the voter model is accurate but slow, while the majority rule is fast but less accurate. 
To be efficient, fast and accurate decision-making mechanisms are needed. 
Researchers frequently engineer new or improve existing decision-making mechanisms to increase efficiency, for example, using  Bayesian hypothesis testing\cite{Shan2020} or the Ising model~\cite{Palina2019}. 
While methods of evolutionary computation have been used to study human collective decision-making from an evolutionary perspective~\cite{10.1162/ARTL_a_00178}, they have only rarely been used to generate collective decision-making mechanisms~\cite{almansoori2021evolution} despite their proven capability to generate controllers for various other swarm behaviors, such as collective motion~\cite{trianni2008evolutionary}. 

\begin{figure}[t]
    \centering
    \subfloat[arena \label{fig:arena}]{\includegraphics[width=0.55\linewidth]{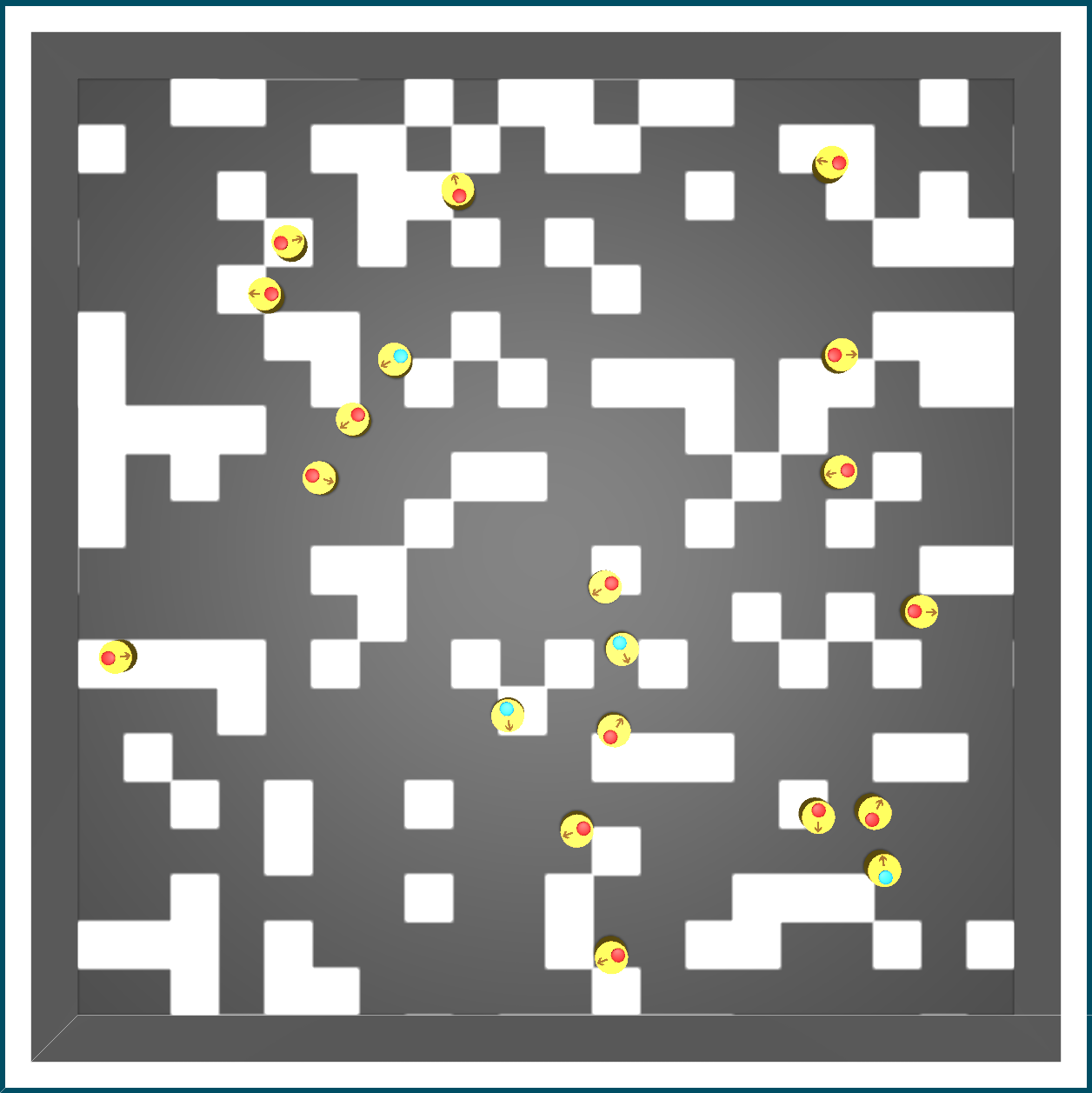}}
    \subfloat[robot \label{fig:robot}]{\includegraphics[width=0.4\linewidth]{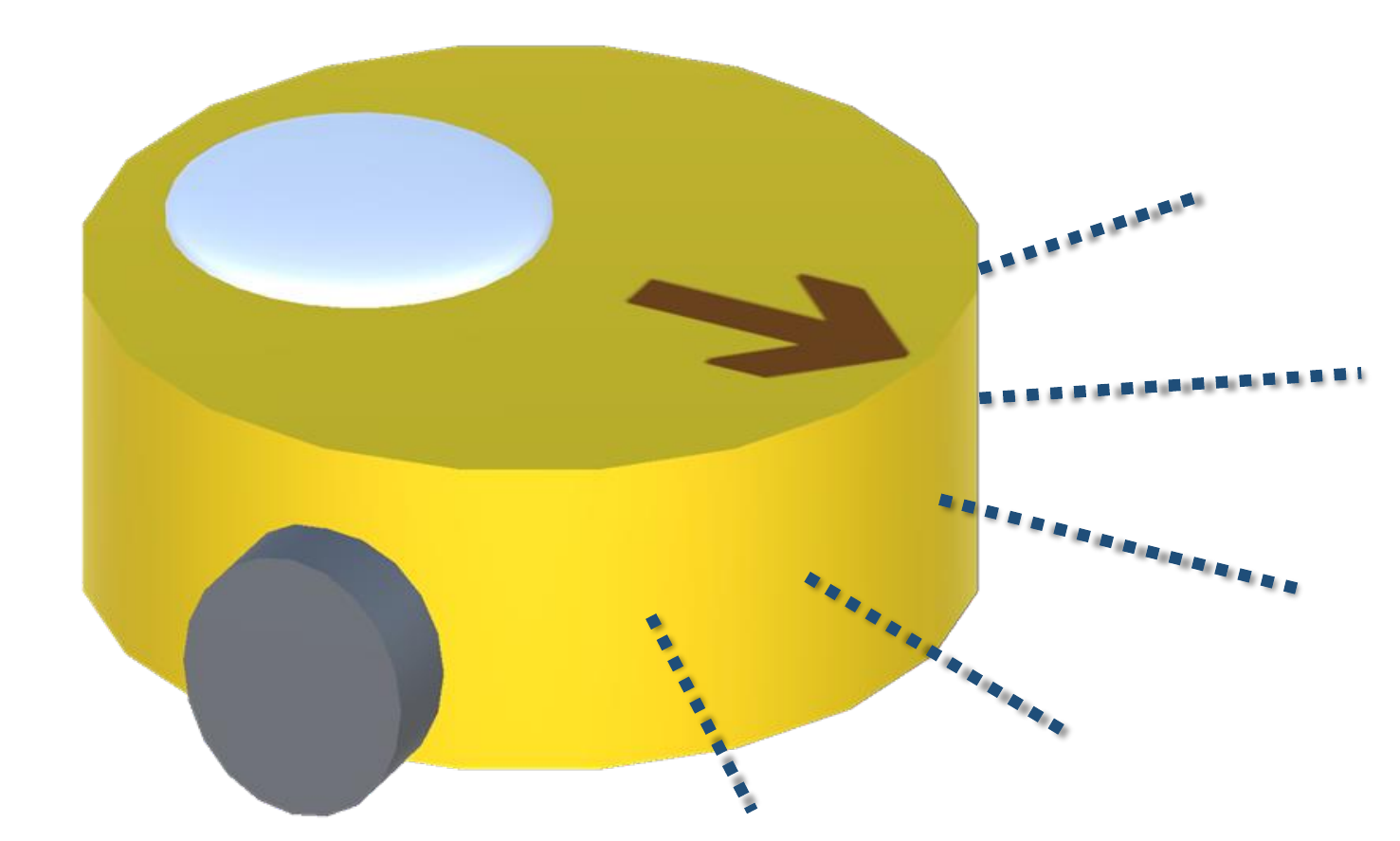}}  
    \caption{Illustration of the experimental setup~(a) with black and white tiles representing the environmental features in the collective perception task and robot model~(b) with five horizontal IR sensors (dashed lines), a ground sensor (not visible), and a LED indicating its current opinion (red representing black, blue representing white).}
\end{figure}
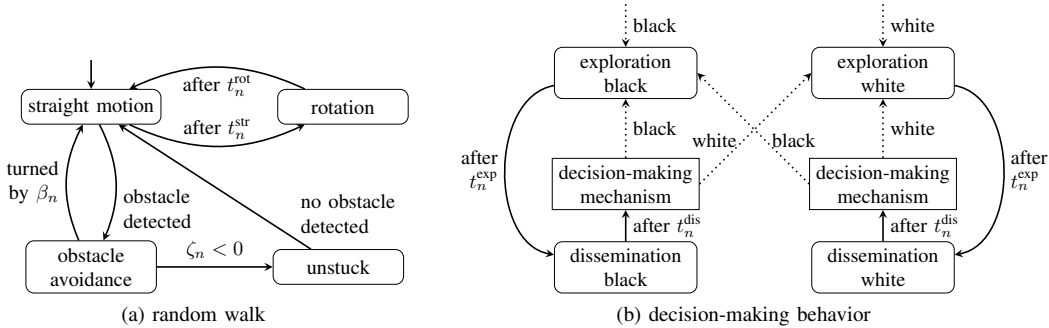
\begin{figure*}[t]
    \centering
    \subfloat[random walk \label{fig:fsm_motion}]{\resizebox{0.31\linewidth}{!}{
		\tikzstyle{arrow} = [thick,->,>=stealth]
			\begin{tikzpicture}[node distance=2cm]
				\node (motion) [rectangle, rounded corners, draw=black, minimum width=2.25cm, minimum height=0.6cm,text centered] {straight motion};
				\node (start) [above=0.5cm of motion] { };
				\node (avoidance) [rectangle, rounded corners, minimum width=2.25cm, minimum height=0.6cm,text centered, draw=black, below=of motion,align=center] {obstacle\\avoidance};
				\node (unstuck) [rectangle, rounded corners, minimum width=2.25cm, minimum height=0.6cm,text centered, draw=black, right=of avoidance] {unstuck};
				\node (rotation) [rectangle, rounded corners, minimum width=2.25cm, minimum height=0.6cm,text centered, draw=black, right=of motion] {rotation};
				
				\draw [arrow] (start) -- (motion);
				\draw [arrow] (avoidance) to [above] node{$\zeta_n < 0$} (unstuck);
				\draw [arrow] (motion) to [bend left=25, below right, align=left] node{obstacle\\detected} (avoidance);
				\draw [arrow] (avoidance) to [bend left=25, left, align=right] node{turned\\by $\beta_n$} (motion);
				\draw [arrow] (motion) to [bend right=25, above,align=center] node{after $t^{\textrm{str}}_n$} (rotation); 
				\draw [arrow] (rotation) to [bend right=25, below,align=center] node{after $t^{\textrm{rot}}_n$} (motion);
				\draw [arrow] (unstuck) to node[below,align=left, xshift=2.3cm]{no obstacle\\detected} (motion);
			\end{tikzpicture}}}
   \hspace{3mm}
    \subfloat[decision-making behavior \label{fig:psfm_decision}]{\resizebox{0.45\linewidth}{!}{
		 		\tikzstyle{arrow} = [thick,->,>=stealth]
	\begin{tikzpicture}[node distance=2cm]
		\node (exploration) [rectangle, rounded corners, minimum width=2.5cm, minimum height=0.6cm,text centered, draw=black, align=center] {exploration\\black};
		\node (start) [above=0.75cm of exploration] { };
		\node (dissemination) [rectangle, rounded corners, minimum width=2.5cm, minimum height=0.6cm,text centered, draw=black, below=2.5cm of exploration, align=center] {dissemination\\black};
		\node (decision) [rectangle, minimum width=2.5cm, minimum height=0.6cm,text centered, draw=black, above=0.55cm of dissemination, align=center] {decision-making\\mechanism};
		\node (exploration2) [rectangle, rounded corners, minimum width=2.5cm, minimum height=0.6cm,text centered, draw=black, align=center, right=of exploration] {exploration\\white};
		\node (start2) [above=0.75cm of exploration2] { };
		\node (dissemination2) [rectangle, rounded corners, minimum width=2.5cm, minimum height=0.6cm,text centered, draw=black, right=of dissemination, align=center] {dissemination\\white};
		\node (decision2) [rectangle, minimum width=2.5cm, minimum height=0.6cm,text centered, draw=black, above=0.55cm of dissemination2, align=center] {decision-making\\mechanism};
		\draw [arrow,dotted] (start) -- node[right,align=left]{black}  (exploration);
		\draw [arrow,dotted] (start2) -- node[right,align=left]{white}  (exploration2);
		\draw [arrow] (exploration) to [bend right=80, left, align=right] node{after\\$t_n^\text{exp}$} (dissemination);
		\draw [arrow] (exploration2) to [bend left=80, right, align=left] node{after\\$t_n^\text{exp}$} (dissemination2);
		\draw [arrow] (dissemination) to [right,align=left] node{after $t_n^\text{dis}$} (decision);
		\draw [arrow] (dissemination2) to [right,align=left] node{after $t_n^\text{dis}$} (decision2);
		\draw [arrow,dotted] (decision2) to [right,align=left] node{white} (exploration2);
		\draw [arrow,dotted] (decision) to [right,align=left] node{black} (exploration);
		\draw [arrow,dotted] (decision.east) to node[left,align=left,yshift=-0.2cm,xshift=-0.2cm]{white}(exploration2.west);
		\draw [arrow,dotted] (decision2.west) to node[right,align=left,yshift=-0.2cm, xshift=0.2cm]{black} (exploration.east);
	\end{tikzpicture}}}
    \caption{Finite state machine for robot motion~(a) and probabilistic finite state machine for decision-making~(b) based on Valentini et al.~\cite{Valentini2016}. Time periods~$t_n^\textrm{rot}$, $t_n^\textrm{str}$, $t_n^\textrm{exp}$, and $t_n^\textrm{dis}$ are randomly sampled. Angle $\beta_n$ is $180^\circ$ plus a randomly sampled value. Buffer values~$\zeta_n$ smaller than zero indicate that the robot mainly rotated and is potentially stuck between obstacles.} 
    \label{fig:fsms}
\end{figure*}
The limited literature includes the work of Morlino et al.~\cite{Morlino2012} who evolved decision-making mechanisms using a task-specific reward to determine the density of an environmental feature but relied on global communication, which contradicts the principles of swarm robotics. 
In addition, Almansoori et al.~\cite{almansoori2021evolution, Almansoori2022} evolved artificial neural networks (ANN) as decision-making behaviors in the collective perception scenario~\cite{Valentini2016} by rewarding swarm members for having the correct opinion in the second half of the evaluation.
That is, they rely on a task-specific fitness function that rewards the collective perception task. 
Almansoori et al. compare their best-evolved decision-making behavior only against the voter model, omitting a comparison against the usually faster majority rule.
The authors neither studied the influence of task-independent rewards nor the scalability with problem difficulty.  

In this paper, we also evolve collective decision-making mechanisms in the collective perception scenario. 
In our approach, the evolved decision-making mechanisms are plugged into a probabilistic finite state machine (PFSM) for direct modulation of positive feedback~\cite{Valentini2016}. 
We study three different fitness functions: (i)~a task-specific fitness function, (ii)~a task-independent fitness function relying on an intrinsic motivation~\cite{Mirolli2013}, and (iii)~a hybrid fitness function combining the two previous. 
The task-specific fitness function rewards the swarm members for having the correct opinion in the last time step of the evaluation and potentially ensures that collective decision-making behaviors emerge.
Our task-independent reward for prediction accuracy could lead to faster decision-making since a collective opinion is easy to predict.
Task-independent rewards and, in particular, intrinsic motivations have only been used rarely in multi-agent and swarm settings~\cite{Khan2018} and, to the best of our knowledge, not yet for collective decision-making tasks. 
These rewards give the evolutionary process the freedom to come up with creative and innovative solutions because the fitness function does not directly constrain how the collective decision-making task is solved.  
However, the emergence of desired behaviors is neither provoked nor guaranteed. 
Our main contribution is that we are, to the best of our knowledge, the first to 
study the influence of task-independent rewards on the evolution of collective decision-making mechanisms for the collective perception task. 
We show that the use of a purely task-independent fitness function requires a careful configuration of the scenario to result in useful swarm behaviors and analyze the impact of the task-independent reward in the hybrid fitness function on the performance of the evolved decision-making mechanisms.      
In addition, we highlight the competitiveness of the evolved decision-making mechanisms to the voter model and the majority rule in benchmark experiments and show that the evolution of decision-making mechanisms for easier problem difficulties can result in better scalability with problem difficulty.  
We introduce our methods in Sec.~\ref{sec:methods} and present our results from the evolutionary runs and the benchmark experiments in Sec.~\ref{sec:results}. 
Sec.~\ref{sec:conclusion} concludes our paper with a summary of the results and a discussion of future work. 
All figures in this paper and supporting materials are available online~\cite{anonymous_2023_7548384}.

\section{Methods} \label{sec:methods}

\subsection{Experimental Setup}

In our simulations, we closely follow Valentini et al. in the setup of the collective perception scenario~\cite{Valentini2016}.

\subsubsection{Arena}

We use a square arena of $2~\textrm{m} \times 2~\textrm{m}$ bounded by walls (see Fig.~\ref{fig:arena}) in the physics-based BeeGround~\cite{Lim2021} simulator.  
The two environmental features for the collective perception task are represented by black and white tiles of $10~\textrm{cm} \times 10~\textrm{cm}$ each on the arena surface.
We vary problem difficulty~$\rho^*$, that is, the ratio of the more frequent feature to the less frequent feature~\cite{Valentini2016}, in our experiments.
Problem difficulty~$\rho^*$ is given by 
\begin{equation}
    \rho^* = \min \left(\frac{\rho_{\textrm{white}}}{\rho_{\textrm{black}}}, \frac{\rho_{\textrm{black}}}{\rho_{\textrm{white}}}\ \right) 
    \, , 
\end{equation}
$\rho_\textrm{white}$ is the percentage of white tiles and $\rho_\textrm{black}$ the percentage of black tiles. 
Thus, an arena with $20~\%$ of black tiles and $80~\%$ of white tiles has a problem difficulty of~$\rho^* = 0.25$ and a problem difficulty~$\rho^*$ of~$1$ means that half the tiles are black and the other half are white. 
BeeGround can simulate real time up to 100~times faster without loosing sampling resolution.
We give all times in real time in this paper.

\subsubsection{Robot} \label{sec:robot}

Our custom robot model (see Fig.~\ref{fig:robot}) is similar to popular robot platforms used in swarm robotics, such as the e-puck~\cite{mondada2009puck}. 
The robot has a diameter of $7~\textrm{cm}$, a differential drive with a maximum speed of $10~\frac{\textrm{cm}}{\textrm{s}}$, and a LED on top indicating its current opinion. 
Its five horizontal frontal proximity sensors have a range of $10~\textrm{cm}$ and are updated every $0.15~\textrm{s}$. 
The robot measures every~$0.2~\textrm{s}$ whether the arena surface below it is black~($0$) or white~($1$) with a binary ground sensor. 
Each robot can broadcast messages containing its ID and current opinion to its neighbors in a $70~\textrm{cm}$ radius (i.e., local communication) and stores the opinions of up to four unique neighbors in a message queue.
As input for our decision-making mechanisms, we aggregate the information stored in the message queue into two virtual sensors. 
Sensor~$s_0$ gives the percentage of neighbor opinions with value white in the message queue and sensor~$s_1$ gives the number of received messages normalized by the maximum message queue length. 
We use a swarm of $N = 20$ robots resulting in a swarm density (i.e., area covered by robots divided by arena size) of approximately~$0.02$ in all experiments.

We base our implementation of the robot behavior (i.e., motion and decision-making) on the common low-level control routines used by Valentini et al.~\cite{Valentini2016} in their initial study on collective perception.
Each robot concurrently executes two finite state machines (see Fig.~\ref{fig:fsms}): a state machine implementing low-level motion routines and a probabilistic finite state machine implementing the decision-making behavior.    

The finite state machine for robot motion implements a random walk behavior, see Fig.~\ref{fig:fsm_motion}. 
Each robot alternates between moving straight for a random time period~$t_n^{\textrm{str}}$ sampled from an exponential distribution with a mean of~$40~\textrm{s}$ and turning on the spot in a random direction for a random time period~$t_n^{\textrm{rot}}$ sampled from a uniform distribution with bounds $[0~\textrm{s}, 4.5~\textrm{s}]$.
When a robot detects an obstacle (i.e., wall or another robot) while moving straight, the robot will turn by $\beta_n = 180^\circ + x_\textrm{rand}$ with $x_\textrm{rand}$ being sampled from a uniform distribution with bounds $[-25^\circ, 25^\circ]$. 
Since thrashing is not prevented by this approach, we introduce the additional state \textit{unstuck} in which the robot rotates in a random direction until no more obstacles are detected. 
The \textit{unstuck} state is activated when a robot has mainly rotated recently and is thus potentially stuck between obstacles. 
To determine if a robot has mainly rotated recently, we introduce buffer value~$\zeta_n$. 
The buffer value decreases when the robot is rotating (i.e., the active state is rotation or obstacle avoidance) and increases up to a maximum of $7.5~\textrm{s}$ when the robot is moving straight (i.e., the 
active state is straight motion). 
Buffer values~$\zeta_n$ lower than zero indicate that the robot has mainly rotated and lead to the activation of the unstuck state. 

The robots' decision-making behavior is implemented as in the direct modulation of decision strategies~\cite{Valentini2016}, see Fig.~\ref{fig:psfm_decision}. 
Different decision-making mechanisms (e.g., voter model~\cite{Valentini2014}, majority rule~\cite{Valentini2015}, evolved decision-making mechanisms) can be applied in this decision-making behavior. 
The PFSM implementing the collective decision-making behavior has four states: an exploration and a dissemination state for each of the two environmental features (i.e., black and white). 
Each robot explores its environment by sampling the local ground color for a time period $t_n^\textrm{exp}$ drawn from an exponential distribution with a mean of~$10~\textrm{s}$.
During this exploration phase, the robot determines a quality estimate~$\hat{\rho}_n$ of its current opinion, that is, the ratio of time it perceived the color associated to its current opinion to the overall duration of the current exploration phase.  
Afterward, the robot broadcasts with $1~\textrm{Hz}$ its current opinion to its local neighborhood for a time period~$t_n^\textrm{send}$ sampled from an exponential distribution with a mean of $T_\textrm{send}\hat{\rho}_n$. 
We set design parameter $T_\textrm{send}$ to $10~\textrm{s}$ in our experiments. 
Thus, robots modulate positive feedback as higher-quality opinions result in longer dissemination phases. 
Subsequent to sending its opinion, the robot records its neighbors' opinions for $t_n^\textrm{receive} = 3~\textrm{s}$. 
After $t_n^\textrm{dis} = t_n^\textrm{send} + t_n^\textrm{receive}$, the robot applies a decision-making mechanism to update its current opinion and switches to the respective exploration state. 

\subsection{Evolution} \label{sec:evo}

\tikzset{%
	every neuron/.style={
		circle,
		draw,
		minimum size=0.7cm
	},
	neuron missing/.style={
		draw=none, 
		scale=2,
		text height=0.333cm,
		execute at begin node=\color{black}$\vdots$
	},
}

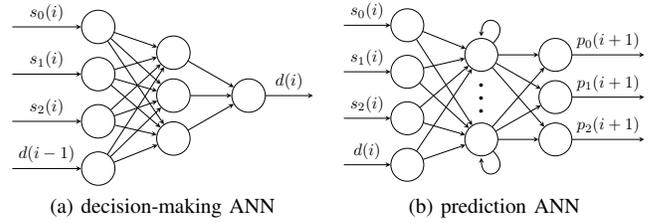
\begin{figure}
	\subfloat[decision-making ANN
	]{
		\resizebox {0.47\linewidth} {!} {
			\begin{tikzpicture}[x=0.9cm, y=0.9cm, >=stealth,]
				\foreach \m/\l [count=\y] in {1,2,3,4}
				\node [every neuron/.try, neuron \m/.try] (input-\m) at (0,2 -\y*1.1) {};
				\foreach \m [count=\y] in {1,2,3}
				\node [every neuron/.try, neuron \m/.try ] (hidden-\m) at (1.75,1.3-\y) {};
				\foreach \m [count=\y] in {1}
				\node [every neuron/.try, neuron \m/.try ] (output-\m) at (3.5,0.3-\y) {};
				\draw [<-] (input-1) -- ++(-2.0,0)
				node [above, midway] {$s_{0}(i)$};
				\draw [<-] (input-2) -- ++(-2.0,0)
				node [above, midway] {$s_{1}(i)$};
				\draw [<-] (input-3) -- ++(-2.0,0)
				node [above, midway] {$s_{2}(i)$};
				\draw [<-] (input-4) -- ++(-2.0,0)
				node [above, midway] {$d(i-1)$};
				\draw [->] (output-1) -- ++(1.5,0)
				node [above, midway] {$d(i)$};
				\foreach \i in {1,2,3,4}
				\foreach \j in {1,2,3}
				\draw [->] (input-\i) -- (hidden-\j);
				\foreach \i in {1,2,3}
				\foreach \j in {1}
				\draw [->] (hidden-\i) -- (output-\j);
		\end{tikzpicture} }\label{fig:actor} } 
	\subfloat[prediction ANN 
	]{
		\resizebox {0.47\linewidth} {!} {
			\begin{tikzpicture}[x=0.9cm, y=0.9cm, >=stealth,]
				\foreach \m/\l [count=\y] in {1,2,3,4}
				\node [every neuron/.try, neuron \m/.try] (input-\m) at (0,2 -\y*1.1) {};
				\foreach \m [count=\y] in {1,missing,2}
				\node [every neuron/.try, neuron \m/.try ] (hidden-\m) at (1.75,1.2-\y) {};
				\foreach \m [count=\y] in {1,2,3}
				\node [every neuron/.try, neuron \m/.try ] (output-\m) at (3.5,1.2-\y) {};
				\draw [<-] (input-1) -- ++(-1.5,0)
				node [above, midway] {$s_{0}(i)$};
				\draw [<-] (input-2) -- ++(-1.5,0)
				node [above, midway] {$s_{1}(i)$};
				\draw [<-] (input-3) -- ++(-1.5,0)
				node [above, midway] {$s_{2}(i)$};
				\draw [<-] (input-4) -- ++(-1.5,0)
				node [above, midway] {$d(i)$};
				\foreach \l [count=\i] in {0, 1}
				\draw [->] (output-\i) -- ++(2.1,0)
				node [above, midway] {$p_{\l}(i+1)$};
				\draw [->] (output-3) -- ++(2.1,0)
				node [above, midway] {$p_{2}(i+1)$};
				\foreach \i in {1,2,3,4}
				\foreach \j in {1,...,2}
				\draw [->] (input-\i) -- (hidden-\j);
				\foreach \i in {1,...,2}
				\foreach \j in {1,2,3}
				\draw [->] (hidden-\i) -- (output-\j);
				\draw[->,shorten >=1pt] (hidden-1) to [out=45,in=90,loop,looseness=5.8] (hidden-1);
				\draw[->,shorten >=1pt] (hidden-2) to [out=315,in=270,loop,looseness=5.8] (hidden-2);
		\end{tikzpicture}  }\label{fig:predictor}
	}
	\caption{ANNs for evolving collective decision-making mechanisms. The decision-making ANN~(a) is used for all of our three different fitness functions, while the prediction ANN~(b) is only used for the two fitness functions with a task-independent component. $s_0(i), \dots, s_2(i)$ are the sensor values at propagation~$i$ and $p_0(i+1),\dots,p_2(i+1)$ are the predictions for the respective sensor values at the next ANN propagation~$i+1$. $d(i-1)$ gives the robot's opinion (i.e., black or white).}
	\label{fig:ANNs}
\end{figure}
\begin{table}
    \centering
    \caption{Parameters}
    \begin{tabular}{ll}
        \textbf{parameter} & \textbf{value} \\ \hline 
         max. generations & 300 ($F_{MS}$), 600 ($F_{TS}$, $F_{HB}$)  \\
         evaluation length &  $200~\textrm{s}$ \\
         evaluations per genome & 6 \\ 
         population size & 50 \\
         parent selection & fitness proportionate \\ 
         survivor selection & age-based \\
         elitism & 1 \\ 
         crossover & none \\
         mutation rate & 0.2 \\ \hline
         swarm size~$N$ & 20 \\ 
         number of sensors~$R$ & 3 \\ 
         problem difficulties (evolution) & $\{0.25, 0.52\}$ \\
         problem difficulties (benchmarks) & $\{0.25, 0.52, 0.67, 0.82 \}$\\ 
    \end{tabular}
    \label{tab:evo}
\end{table}

In our experiments, we evolve three-layered ANNs as collective decision-making mechanisms (see Fig.~\ref{fig:actor}) that are plugged into the PFSM (see Fig.~\ref{fig:psfm_decision}). 
This feed-forward ANN receives four inputs: the percentage of neighbor opinions with value white~$s_0(i)$, the number of received messages normalized by the maximum message queue length~$s_1(i)$, the current ground sensor reading~$s_2(i)$ at ANN propagation~$i$, and the robot's last opinion~$d(i-1)$. 
The ANN outputs the robot's updated opinion~$d(i) \in \{\textrm{white}, \textrm{black}\}$. 
Inputs are propagated through the ANN pair of each robot at the end of the dissemination phase. 
Thus, the number of ANN propagations may vary between robots as the time periods determining the length of the exploration and dissemination phases are randomly sampled. 

\begin{figure*}[t]
    \centering
    \subfloat[$F_{TS}$\label{fig:fitness_fts}]{\includegraphics[width=0.32\linewidth]{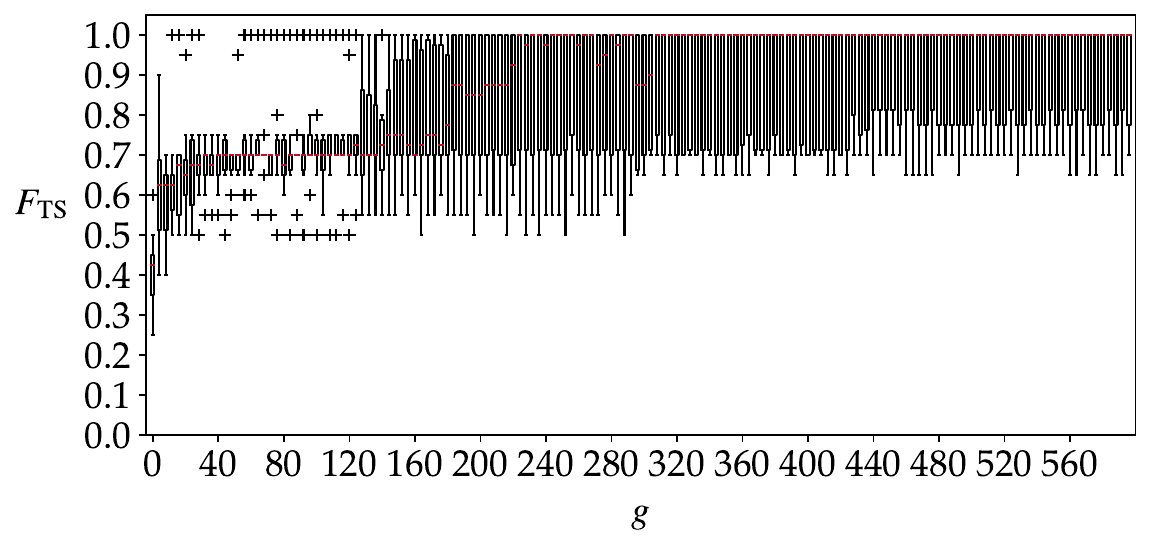}}
    \subfloat[$F_{MS}$\label{fig:fitness_fms}]{\includegraphics[width=0.32\linewidth]{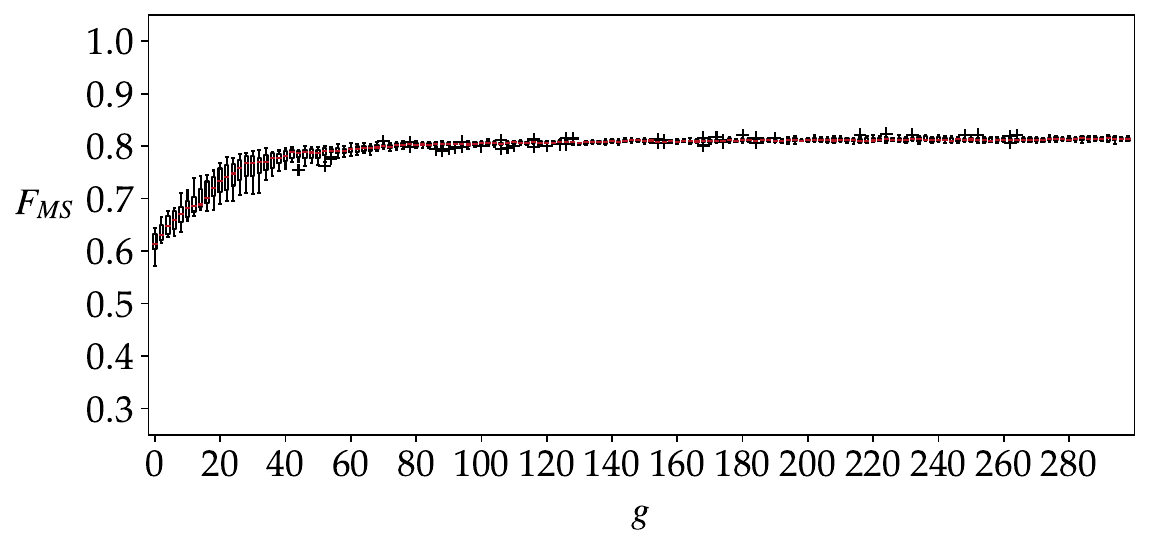}}
    \subfloat[$F_{HB}$\label{fig:fitness_fhb}]{\includegraphics[width=0.32\linewidth]{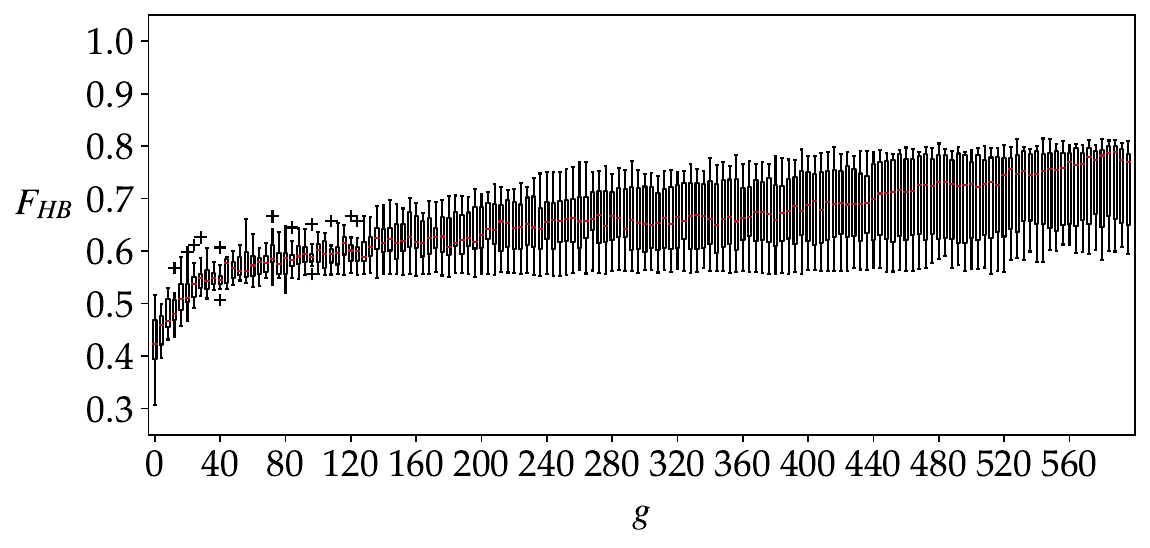}}
    \caption{Best fitness over generations~$g$ of 10 independent evolutionary runs per fitness function in problem difficulty~$\rho^* = 0.52$ with task-specific fitness function~$F_{TS}$ (Eq.~\ref{equ:TS}) rewarding correct opinions in the last time step of an evaluation, task-independent fitness function~$F_{MS}$ (Eq.~\ref{equ:MS}) rewarding prediction accuracy, and hybrid fitness function~$F_{HB}$ (Eq.~\ref{equ:HB}) combining the two previous. For clearer illustration, we only plot the data of every second generation for $F_{MS}$ and for every fourth generation for~$F_{TS}$ and $F_{HB}$.}
    \label{fig:fitmess}

    \centering
    \subfloat[$F_{TS}$\label{fig:dmp_fts}]{\includegraphics[width=0.32\linewidth]{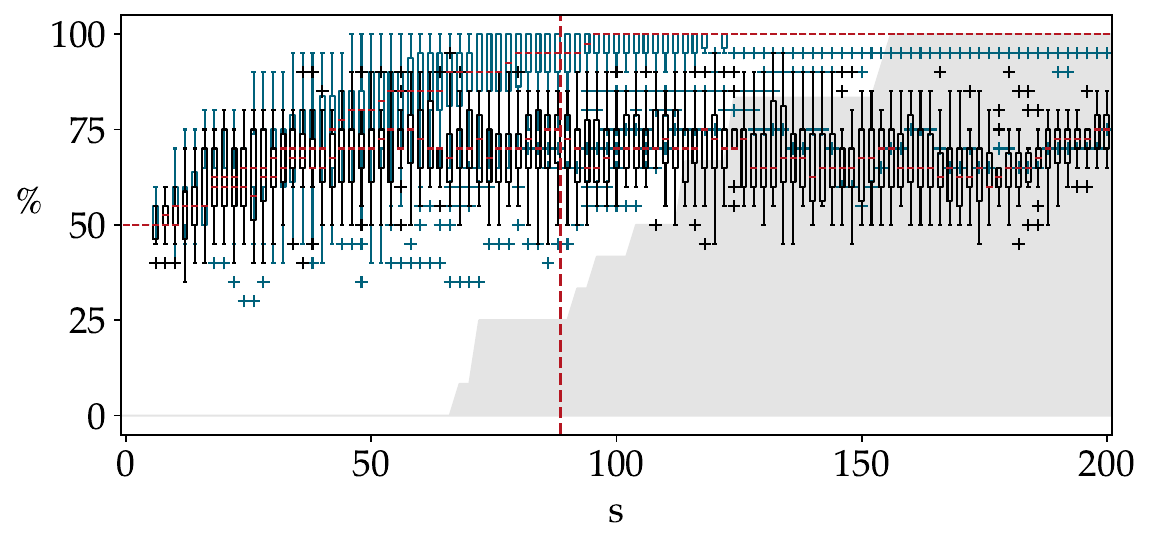}}
    \subfloat[$F_{MS}$\label{fig:dmp_fms}]{\includegraphics[width=0.32\linewidth]{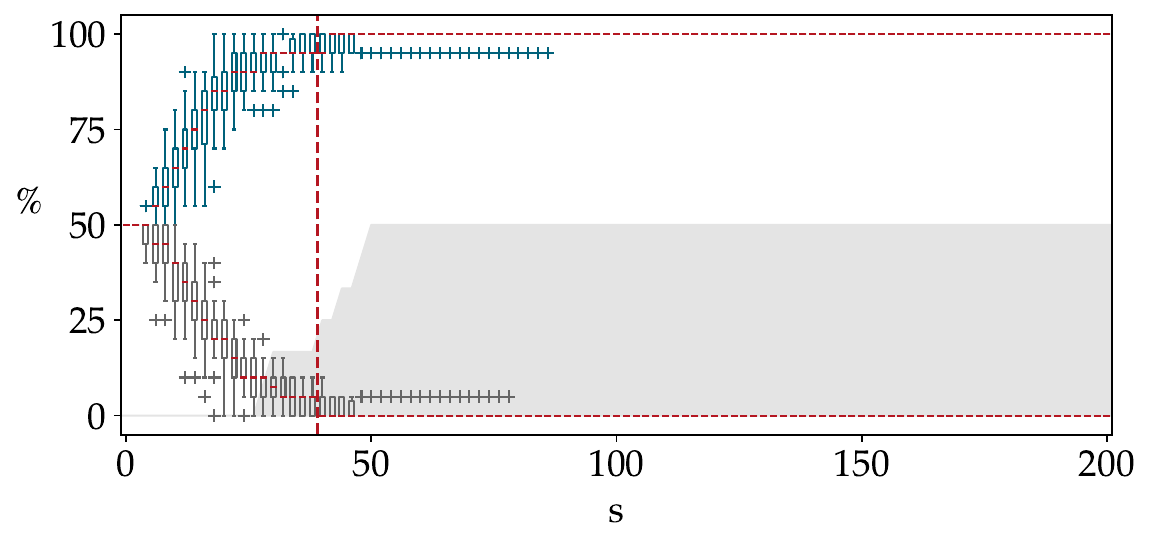}}
    \subfloat[$F_{HB}$\label{fig:dmp_fhb}]{\includegraphics[width=0.32\linewidth]{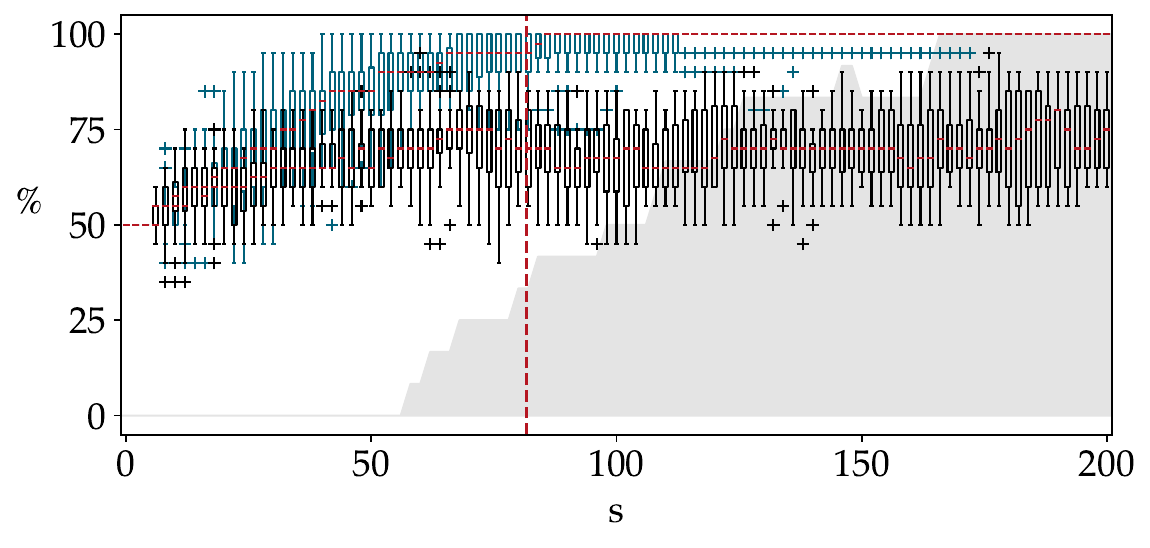}}
    \caption{Decision-making process over seconds~$s$ of the 10 best evolved individuals per fitness function in problem difficulty~$\rho^* = 0.52$ with task-specific fitness function~$F_{TS}$ (Eq.~\ref{equ:TS}), task-independent fitness function~$F_{MS}$ (Eq.~\ref{equ:MS}), and hybrid fitness function~$F_{HB}$ (Eq.~\ref{equ:HB}).
    Boxes give the percentage of swarm members with the more frequent feature as opinion. Blue boxes represent evaluations leading to a consensus for the more frequent feature, black boxes represent evaluations leading to no consensus, and gray boxes represent evaluations leading to a consensus for the less frequent feature. The gray areas give the median exit probability~$E_N$ over time. The median mean consensus times~$\overline{T}_N$ are given by the dashed red lines. Medians are given by the red bars in the boxes. For clearer illustration, we only plot the data of every second time step.  
    }
    \label{fig:dmprocess}
\end{figure*}

We use three different fitness functions to evolve collective decision-making mechanisms: (i)~a task-specific fitness function~$F_{TS}$, (ii)~a task-independent fitness function~$F_{MS}$, and (iii)~a hybrid fitness function~$F_{HB}$ combining the two previous. 
Our task-specific fitness function~$F_{TS}$ rewards higher percentages of swarm members with the correct opinion in the last time step of the evaluation. 
Fitness~$F_{TS}$ is given by 
\begin{equation}
\begin{aligned}
   \label{equ:TS}
    F_{TS} & =  \frac{n_{\textrm{best}}(T-1)}{N} \, ,   
\end{aligned}
\end{equation}
\(n_{\textrm{best}}(T-1)\) is the number of robots in the swarm with the best feature as opinion last time step of an evaluation.

We use our minimize surprise approach as our task-independent fitness function~$F_{MS}$~\cite{kaiser2022}. 
Here, each robot has a prediction ANN (see Fig.~\ref{fig:predictor}) next to the decision-making ANN. 
The prediction ANN is a three-layer recurrent ANN that outputs predictions for the robot's sensor values at the next ANN propagation~$i+1$. 
We reward the prediction accuracy of the predictor ANN, that is, we minimize the prediction error (i.e., surprise). 
Fitness~$F_{MS}$ is defined as  
\begin{equation}
    F_{MS} = \frac{1}{RI} \sum_{n=0}^{N-1} \sum_{i=0}^{i_n-1} \sum_{r=0}^{R-1} (1 - |p^r_n(i) - s^r_n(i)|)
        \label{equ:MS}
\end{equation}
with swarm size~$N$, number of sensors per swarm member~$R$, quantity of ANN propagations~$i_n$ of robot~$n$, total quantity of ANN propagations $I = \sum_{n=0}^{N-1} i_n$ of the swarm, robot~$n$’s prediction $p^r_n(i)$ for the actual value of sensor~$r$ at ANN propagation~$i$ and the actual value $s^r_n(i)$ of sensor~$r$ at ANN propagation~$i$.
Decision-making ANN and prediction ANN are evolved in pairs, that means, only the prediction ANN is directly rewarded but both networks are subject to selection and variation. 
Since we do not explicitly reward collective decision-making in this case, the emergence of desired behaviors is neither provoked nor guaranteed. 

For our hybrid fitness function, we combine the two previous rewards. 
As in the task-independent approach, we evolve pairs of decision-making and prediction ANNs. 
Fitness~$F_{HB}$ is
\begin{equation}
    F_{HB} = \left(\frac{1-F_{TS}}{\kappa}+F_{TS}\right) F_{MS}
    \label{equ:HB}
\end{equation}
with penalty factor $\kappa > 1$, our task-specific fitness function~\(F_{TS}\) (Eq.~\ref{equ:TS}), and our task-independent fitness function~\(F_{MS}\) (Eq.~\ref{equ:MS}). 
Thus, we reduce the reward for prediction accuracy if the swarm did not reach a consensus for the more frequent feature in the last time step of an evaluation.  
The maximum penalty of~$1 - \frac{1}{\kappa}$ is applied if the swarm reaches a consensus for the less frequent environmental feature. 
In our experiments, we use a penalty factor~$\kappa$ of~$2$. 

We evolve the ANNs using a simple evolutionary algorithm.
Candidate solutions encode the synaptic weights of the ANNs and we randomly generate the initial population. 
The parameters for the evolutionary runs are given in Tab.~\ref{tab:evo}.
Each genome is evaluated in six independent evaluations. 
We evaluate the ANNs in three different black-and-white patterns and the inverse of these patterns to avoid bias towards one feature. 
The fitness of a candidate solution is the minimum fitness observed in its six evaluations. 
We do ten independent evolutionary runs per fitness function and problem difficulty~$\rho^* \in \{0.25,0.52\}$.

\subsection{Evaluation Metrics}\label{sec:metrics}

We evaluate the collective decision-making mechanisms using two common metrics: mean consensus time~$\overline{T}_N$ measuring decision speed and exit probability~$E_N$ indicating decision accuracy. 
The consensus time~$T_N$ is the time a swarm needs to reach a first consensus and the mean consensus time~$\overline{T}_N$ gives the average consensus time over all runs leading to a consensus. 
Exit probability~$E_N$ is the percentage of runs in which the swarm successfully reached a consensus for the more frequent feature. 
Since we do 10~independent evolutionary runs per evolutionary setting, we quantify the performance of the evolved decision-making mechanisms using the median value of the ten evaluated best-evolved individuals each for the mean consensus time~$\overline{T}_N$ and the exit probability~$E_N$. 

\subsection{Benchmarks} \label{sec:benchmarks}

We compare the mean consensus time~$\overline{T}_N$ and the exit probability~$E_N$ of the best-evolved decision-making mechanisms of all ten runs per evolutionary setting to the voter model and the majority rule in benchmark experiments to evaluate their competitiveness. 
We run each decision-making mechanism $1\,000$~times for $400~\textrm{s}$ per problem difficulty $\rho^* \in \{0.25,0.52,0.67,0.82\}$.
Consequently, we also test the evolved decision-making mechanisms in problem difficulties for which they have not been optimized to determine their scalability with problem difficulty. 
To ensure comparability, we use the same evaluation settings (i.e., initial robot poses and opinions, and ground patterns of black and white tiles) for the different collective decision-making mechanisms. 
Since we obtained comparable results in control experiments with white or black as the more frequent feature, we set black as the more frequent feature in all settings without loss of generality.  

\section{Results} \label{sec:results}

\begin{figure}
    \centering
     \subfloat[more frequent feature: white]{\includegraphics[width=0.75\linewidth]{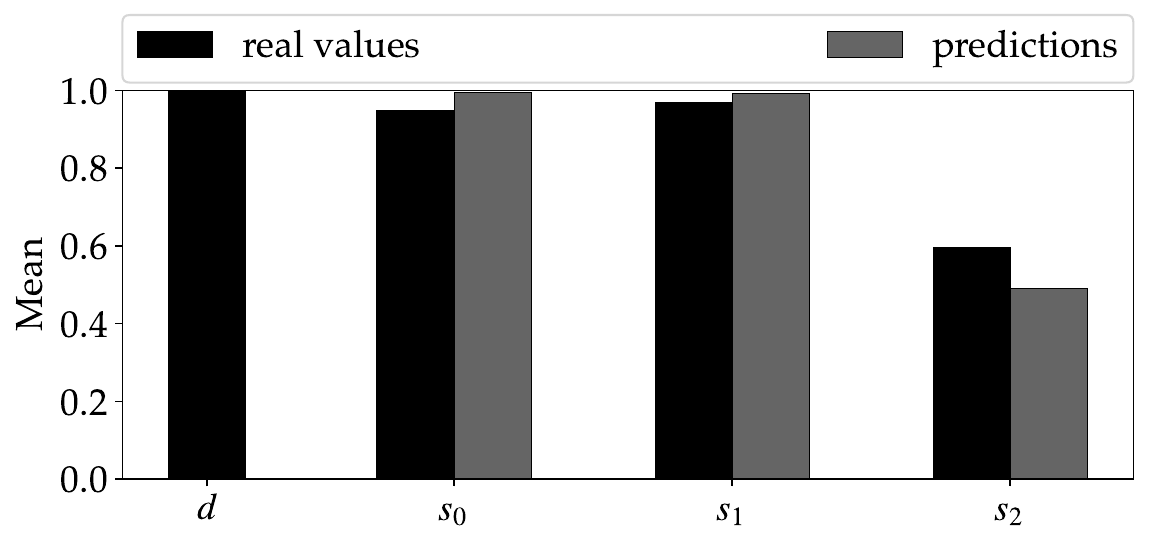}} \\
    \subfloat[more frequent feature: black]{\includegraphics[width=0.75\linewidth]{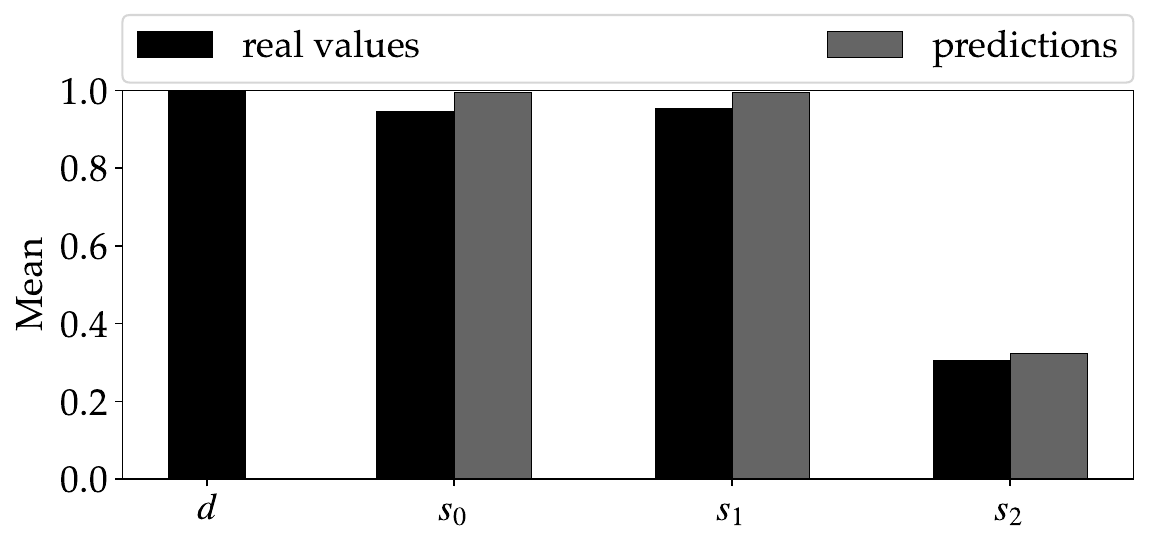}}
    \caption{Mean decision-making ANN output~$d$ and mean inputs to and outputs of the prediction ANN for one representative best-evolved individual using fitness function~$F_{MS}$ with white~(a) and black~(b) as the most frequent environmental feature. 
    Sensor $s_0$ gives the percentage of neighbor opinions with value white, $s_1$ gives the number of received messages normalized by the maximum message queue length, and $s_2$ is the ground sensor reading, see Sec.~\ref{sec:robot}.
    }
    \label{fig:sensor_predictions}
\end{figure}

First, we analyze the success of evolving decision-making mechanisms using our three different fitness functions. 
Afterward, we show the competitiveness of our evolved decision-making mechanisms in benchmark experiments. 

\subsection{Evolution of Decision-Making Mechanisms}

\subsubsection{Task-Specific Fitness Function~$F_{TS}$}

In both problem difficulties (i.e., $\rho^* \in \{0.25,0.52\}$), we find a median best task fitness of~$1.0$ in the last generation of the evolutionary run using our task-specific fitness function~$F_{TS}$ (Eq.~\ref{equ:TS}) that rewards high numbers of swarm members with the correct opinion in the last time step of an evaluation. 
Fig.~\ref{fig:fitness_fts} illustrates best fitness over generations for the 10~independent evolutionary runs in problem difficulty~$\rho^* = 0.52$. 
Compared to the evolutionary runs in problem difficulty~$\rho^* = 0.25$ (see~\cite{anonymous_2023_7548384} for data), we find high variability in fitness in the higher generations in the runs in problem difficulty~$\rho^* = 0.52$. 
This is reflected by the fact that the best evolved decision-making ANNs in all ten runs in problem difficulty~$\rho^* = 0.25$ reach the correct consensus (i.e., median exit probability~$E_N = 100~\%$ and fitness of~$1.0$) in a median mean consensus time~$\overline{T}_N$ of $54.1~\textrm{s}$. 
But only seven out of ten best-evolved individuals in problem difficulty~$\rho^* = 0.52$ reach the correct consensus (i.e., a fitness of~$1.0$) in all of their six evaluations in the last generation. 
Still, we find a median exit probability~$E_N$ of $100~\%$ and a median mean consensus time~$\overline{T}_N$ of $90.7~\textrm{s}$ here, see Fig.~\ref{fig:dmp_fts}. 
Thus, evolution successfully optimizes the decision-making ANNs to make correct collective decisions but harder problem difficulties complicate the optimization of the decision-making mechanisms.

\subsubsection{Task-Independent Fitness Function~$F_{MS}$}

Using our task-independent fitness function~$F_{MS}$ (Eq.~\ref{equ:MS}) rewarding prediction accuracy in problem difficulty~$\rho^* = 0.52$, we find a median best prediction fitness of~$0.81$ in the last generation of the evolutionary run, a median mean consensus time~$\overline{T}_N$ of $41.1~\textrm{s}$, and a median exit probability~$E_N$ of~$50~\%$. 
Thus, evolution successfully optimizes the decision-making and prediction ANN pairs to reach high fitness, see Fig.~\ref{fig:fitness_fms}.
All evaluations lead to a consensus (see Fig.~\ref{fig:dmp_fms}) and decision speed is high (i.e., low mean consensus time), but only half of the evaluations result in a consensus for the more frequent environmental feature (i.e., $E_N = 50~\%$). 
As explained in Sec.~\ref{sec:evo}, each candidate solution is evaluated in six independent evaluations, three of which have black and three of which have white as the most frequent feature. 
The best-evolved individuals lead to a consensus for one of the two options regardless of the actual most frequent feature, that is, a part of the best-evolved individuals always leads to a consensus for black and the other part always leads to a consensus for white, see Fig.~\ref{fig:sensor_predictions}. 
Consequently, each best-evolved individual reaches the correct consensus only by chance in half of its evaluations. 
This simplifies the prediction task because predicting sensor value~$s_0$ (i.e., percentage of neighbors with opinion white) becomes trivial since each swarm member's decision-making ANN outputs the same fixed opinion. 
There is no punishment in the evolutionary process for reaching the wrong consensus, since robot opinions are independent of the actual state of the environment (i.e., the frequency of features here) and only prediction accuracy is rewarded. 
In addition, we find that sensor value~$s_1$ (i.e., the number of received neighbor opinions normalized by the maximum message queue length) is almost always at the maximum of~$1.0$ and thus easy to predict. 
Only the ground sensor value~$s_2$ depends on the actual more frequent feature and the predictions are optimized accordingly in the evolutionary process. 
In this way, evolution exploits the easiest possible way to optimize fitness but does not result in collective decision-making mechanisms. 
Based on these results, we refrain from evolving decision-making ANNs in problem difficulty~$\rho^* = 0.25$ and exclude the best-evolved individuals from this setting from the benchmark experiments.

\subsubsection{Hybrid Fitness Function~$F_{HB}$}

We find median best fitnesses of~$0.88$ and~$0.77$ in the last generation of the evolutionary runs for problem difficulties~$0.25$ and~$0.52$, respectively, using our hybrid fitness function~$F_{HB}$ (Eq.~\ref{equ:HB}) that combines the two previous rewards.  
As for the evolutionary runs with our task-specific fitness function~$F_{TS}$, we find more variability in fitness in the later generations of the run in the setting with the harder problem difficulty (i.e., $\rho^* = 0.52$), see Fig.~\ref{fig:fitness_fhb}. 
This has probably two related reasons: (i)~the problem difficulty makes reaching a consensus hard and (ii)~the hybrid fitness function rigorously punishes ANN pairs that do not lead to the correct consensus or a consensus at all.
The best-evolved individuals in problem difficulty~$\rho^* = 0.25$ reach a consensus for the more frequent feature in all evaluations (i.e., $E_N = 100\%$).
We find a median mean consensus time $\overline{T}_N$ of $56.7~\textrm{s}$ here.
Although we find a median exit probability~$E_N$ of $100~\%$ and a median mean consensus time~$\overline{T}_N$ of $83.7~\textrm{s}$ for the ten best-evolved individuals in the harder problem difficulty (i.e., $\rho^* = 0.52$), not all evaluations lead to a consensus, see Fig.~\ref{fig:dmp_fhb}. 
We find that two of the ten best-evolved individuals do not lead to consensus in any of their six evaluations, and one best-evolved individual leads to consensus in only two of its six evaluations. 
There is however a tendency towards correct consensus in all runs in which the swarm has not reached consensus. 
Increasing the number of generations and the evaluation length for the harder problem difficulty would probably lead to less variability and better convergence of the fitness curve as well as to the generation of decision-making mechanisms in every evolutionary run.
Overall, we successfully evolved collective decision-making mechanisms using our hybrid fitness function.

\subsection{Benchmarks}

\begin{table*}[t]
	\caption{Mean consensus times~$\overline{T}_N$ and exit probabilities~$E_N$ for the benchmark runs with voter model (VM), majority rule (MR), evolved decision-making mechanisms using our task-specific fitness function in problem difficulties $0.25$ (TS-E) and $0.52$ (TS-H), and our hybrid fitness function in problem difficulties $0.25$ (HB-E) and $0.52$ (HB-H). Values are calculated based on~$1\,000$~runs of~$400~\text{s}$ per decision-making mechanism and problem difficulty~$\rho^*$. 
 For the evolved decision-making mechanisms, the median values of the respective 10 best-evolved individuals are given. The best value per problem difficulty is marked in bold.
\label{tab:metrics}
}
\centering
		\begin{tabular}{rrrrrrrrr}
		\hline 
		{decision-making}
		& \multicolumn{2}{c}{$\rho^* =0.25$} & \multicolumn{2}{c}{$\rho^* =0.52$} & \multicolumn{2}{c}{$\rho^* =0.67$} & \multicolumn{2}{c}{$\rho^* =0.82$} \\
		mechanism & & & & & & & & 
		\\ 
		 & \multicolumn{1}{c}{$\overline{T}_N$} &  \multicolumn{1}{c}{$E_N$}&  \multicolumn{1}{c}{$\overline{T}_N$} &  \multicolumn{1}{c}{$E_N$} & \multicolumn{1}{c}{$\overline{T}_N$} &  \multicolumn{1}{c}{$E_N$} &  \multicolumn{1}{c}{$\overline{T}_N$} & \multicolumn{1}{c}{$E_N$}\\ 
	   \hline 
		VM & $94.9~\text{s}$ & \textbf{$\bm{100.0}$~\%} & $154.7~\text{s}$ & $97.1~\text{\%}$ & $192.5~\text{s}$ & $83.4~\text{\%}$  &  $206.3~\text{s}$ & $60.2~\text{\%}$ \\ 
		MR & $69.5~\text{s}$ & $96.7~\text{\%}$ & \textbf{$\bm{84.5}~\text{s}$} & $83.5~\text{\%}$ & \textbf{$\bm{86.8}~\text{s}$} & $72.8~\text{\%}$  &  \textbf{$\bm{94.1}~\text{s}$} & $61.1~\text{\%}$ \\[0.3em] 
		TS-E & \textbf{$\bm{55.7}~\text{s}$} & \textbf{$\bm{100.0}~\text{\%}$} & $85.1~\text{s}$ & $99.6~\text{\%}$ & $111.5~\text{s}$ & \textbf{$\bm{98}~\text{\%}$}  &  $140.1~\text{s}$ & \textbf{$\bm{85.8}~\text{\%}$} \\ 
		TS-H & $57.5~\text{s}$ & \textbf{$\bm{100.0}~\text{\%}$} & $90.7~\text{s}$ & $\bm{99.7}~\text{\%}$ & $123.2~\text{s}$ & $96.2~\text{\%}$  &  $151.1~\text{s}$ & $80.7~\text{\%}$ \\ [0.3em]
  		HB-E & $61.1~\text{s}$ & \textbf{$\bm{100.0}~\text{\%}$} & $107.6~\text{s}$ & $99.4~\text{\%}$ & $140.7~\text{s}$ & $93.2~\text{\%}$  &  $152.0~\text{s}$ & $62.3~\text{\%}$ \\ 
		HB-H & $56.9~\text{s}$ & \textbf{$\bm{100.0}~\text{\%}$} & $93.1~\text{s}$ & $\bm{99.7}~\text{\%}$ & $128.2~\text{s}$ & $96.3~\text{\%}$  &  $166.3~\text{s}$ & $70.6~\text{\%}$ \\ 
		\hline 
	\end{tabular}
\end{table*}
\begin{figure*}
    \centering
     \subfloat[$\rho^* = 0.25$\label{fig:ep025}]{\includegraphics[width=0.25\linewidth]{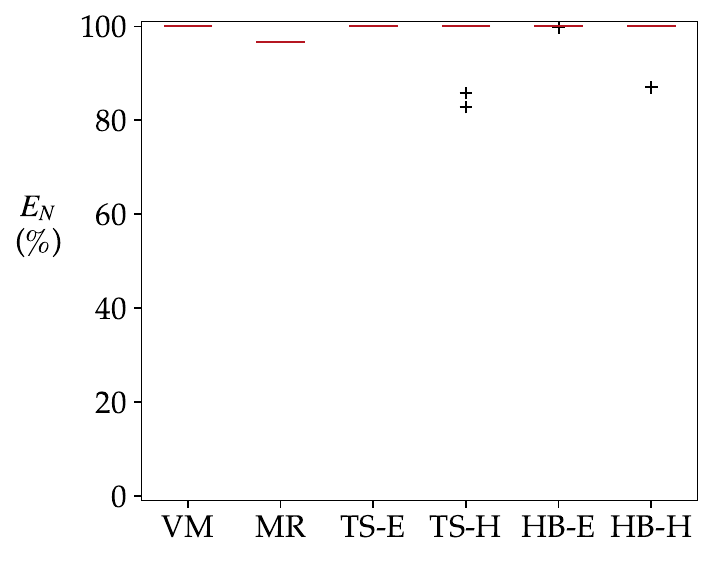}}
    \subfloat[$\rho^* = 0.52$\label{fig:ep052}]{\includegraphics[width=0.25\linewidth]{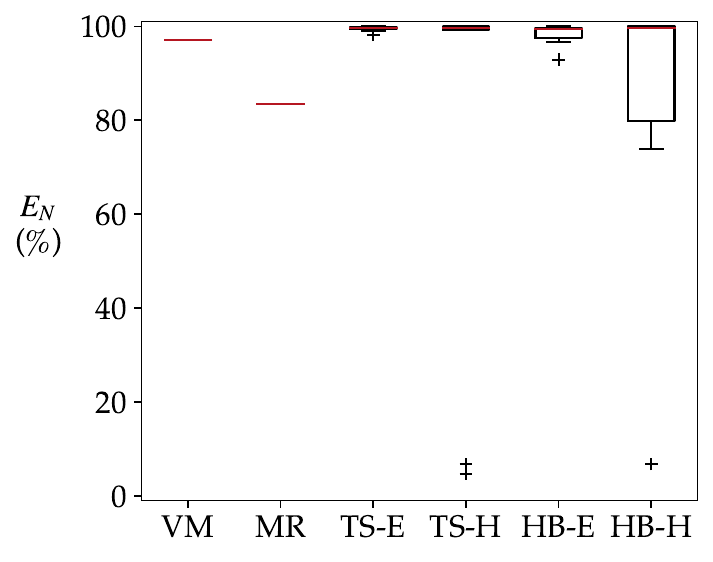}}
    \subfloat[$\rho^* = 0.67$\label{fig:ep067}]{\includegraphics[width=0.25\linewidth]{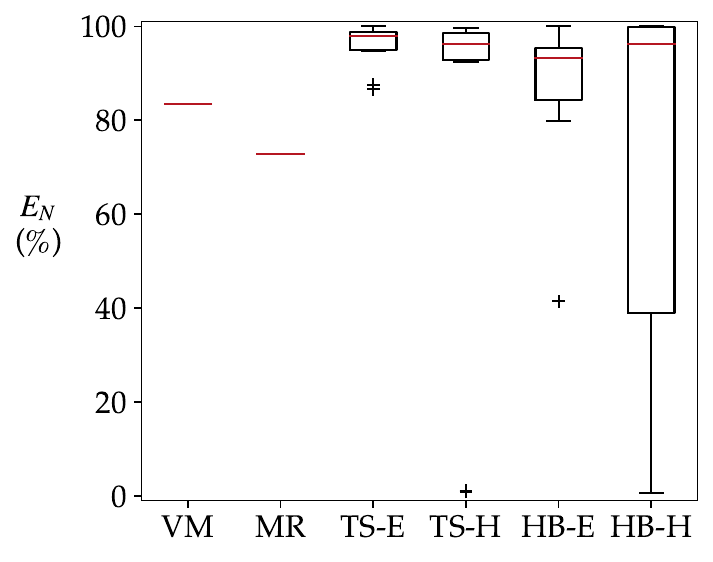}}
    \subfloat[$\rho^* = 0.82$\label{fig:ep082}]{\includegraphics[width=0.25\linewidth]{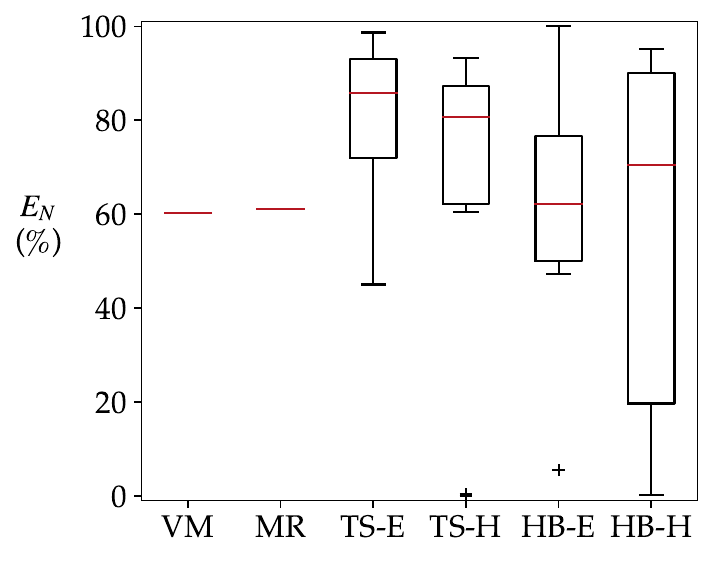}}
    \caption{Exit probabilities~$E_N$ (higher is better) for the benchmark runs with voter model (VM), majority rule (MR), evolved decision-making mechanisms using our task-specific fitness function in problem difficulties of either $0.25$ (TS-E) or $0.52$ (TS-H), and our hybrid fitness function in problem difficulties of either $0.25$ (HB-E) or $0.52$ (HB-H). For the voter model and the majority rule, the plots give one value. For the evolved decision-making mechanisms, the values of the 10 best-evolved individuals are included.}
    \label{fig:exitprobability}
    \subfloat[$\rho^* = 0.25$\label{fig:ct025}]{\includegraphics[width=0.25\linewidth]{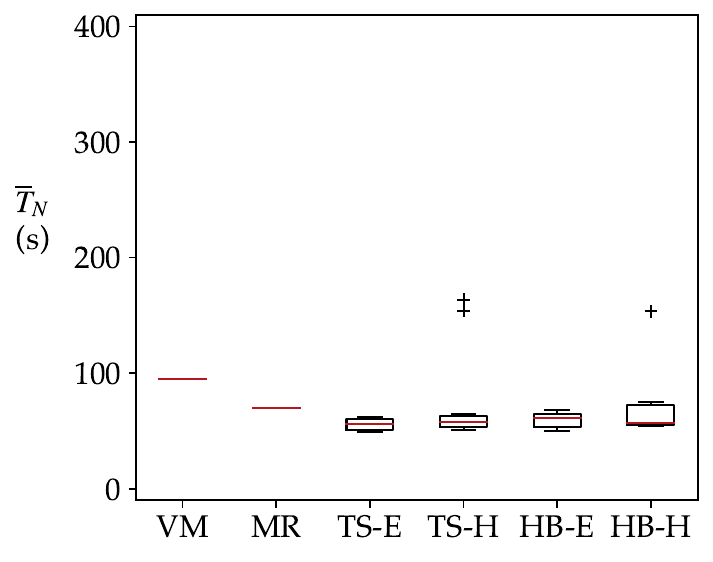}}
    \subfloat[$\rho^* = 0.52$\label{fig:ct052}]{\includegraphics[width=0.25\linewidth]{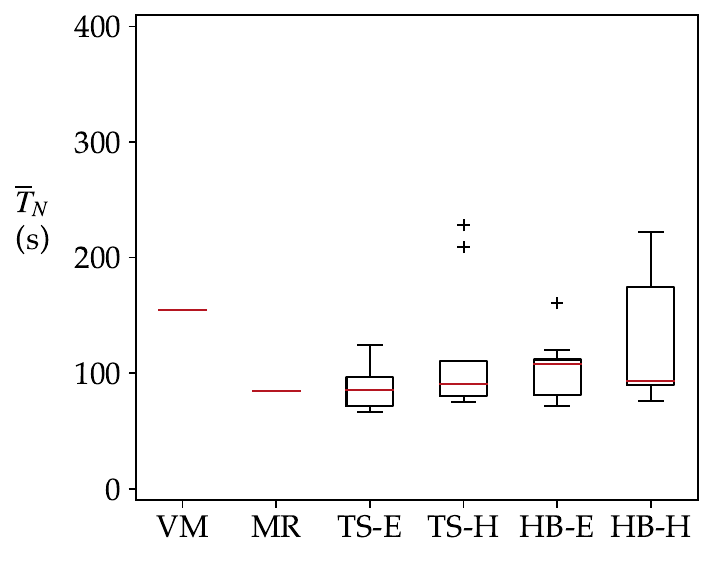}}
    \subfloat[$\rho^* = 0.67$\label{fig:ct067}]{\includegraphics[width=0.25\linewidth]{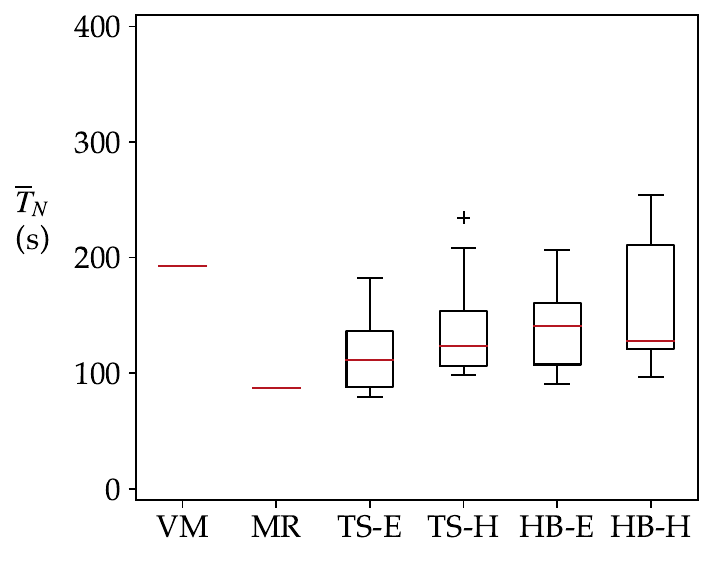}}
    \subfloat[$\rho^* = 0.82$\label{fig:ct082}]{\includegraphics[width=0.25\linewidth]{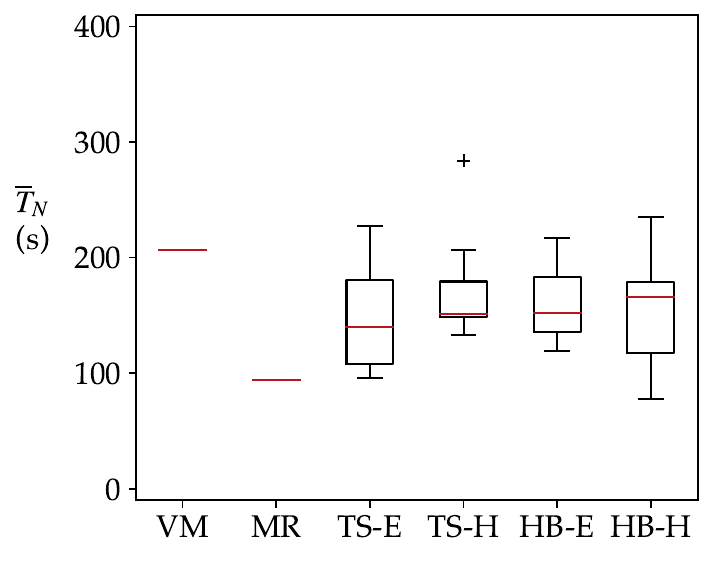}}
    \caption{Mean consensus times~$\overline{T}_N$ (lower is better) for the benchmark runs with voter model (VM), majority rule (MR), evolved decision-making mechanisms using our task-specific fitness function in problem difficulties of either $0.25$ (TS-E) or $0.52$ (TS-H), and our hybrid fitness function in problem difficulties of either $0.25$ (HB-E) or $0.52$ (HB-H). For the voter model and the majority rule, the plots give one value. For the evolved decision-making mechanisms, the values of the 10 best-evolved individuals are included. }
    \label{fig:consensustime}
\end{figure*}

In our benchmark experiments, we evaluate the competitiveness of the evolved decision-making mechanisms to the standard decision-making mechanisms voter model and majority rule (see Sec.~\ref{sec:benchmarks}). 
We only compare against the best-evolved individuals of the evolutionary runs using task-specific fitness function~$F_{TS}$ and hybrid fitness function~$F_{HB}$, since no decision-making mechanisms emerged in the runs using task-independent fitness function~$F_{MS}$. 

For all decision-making mechanisms, we find increasing mean times to consensus~$\overline{T}_N$ and decreasing exit probabilities~$E_N$ with increasing problem difficulty~$\rho^*$, see Tab.~\ref{tab:metrics}. 
As already known for the two decision-making mechanisms, the voter model is more accurate (i.e., higher $E_N$) while the majority rule is faster (i.e., shorter $\overline{T}_N$) in our experiments.
This result is in line with the speed versus accuracy trade-off~\cite{Valentini2015}. 
In the experiments with problem difficulty $\rho^* = 0.82$, that is, the hardest problem difficulty used in our benchmarks, we find a slightly higher exit probability~$E_N$ for the majority rule than for the voter model. 
While the majority rule always leads to a consensus -- either a correct or an incorrect one -- the voter model did not lead to a consensus in $27~\%$ of the runs here. 
We expect that the exit probability~$E_N$ is also higher for the voter model than for the majority rule in this problem difficulty when we would increase runtime at the expense of an even higher mean consensus time~$\overline{T}_N$.

We find equal or higher median exit probabilities~$E_N$ for all evolved decision-making mechanisms compared to the voter model and the majority rule, see Fig.~\ref{fig:exitprobability}. 
Thus, evolution leads to higher decision accuracy. 
In addition, the evolved decision-making mechanisms have lower mean times to consensus~$\overline{T}_N$ than the voter model, that is, they outperform the voter model in decision speed, see Fig.~\ref{fig:consensustime}. 
Compared to the majority rule, the evolved decision-making mechanisms have comparable or slightly higher mean times to consensus~$\overline{T}_N$. 
Overall, the evolved decision-making mechanisms are at least competitive if not more performant than the voter model and the majority rule. 
In contrast to the voter model and the majority rule, the evolved decision-making ANNs receive also the ground sensor values as input for making a decision. 
We expect that this extra input has a beneficial influence on the efficiency of the decision-making mechanisms since the decision does not only depend on the opinions of the robot's neighbor in this case. 

Of all evolved decision-making mechanisms, the best-evolved individuals of the evolutionary runs in problem difficulty~$\rho^*=0.25$ using task-specific fitness function~$F_{TS}$ have the best exit probabilities~$E_N$ and mean times to consensus~$\overline{T}_N$. 
These decision-making mechanisms scale up well in problem difficulty and have even shorter mean times to consensus~$\overline{T}_N$ than the best-evolved individuals in problem difficulty~$\rho^*=0.52$. 
While all ten best-evolved individuals using task-specific fitness function~$F_{TS}$ in the easier problem difficulty work well in all tested problem difficulties, three of the ten best-evolved individuals using task-specific fitness function~$F_{TS}$ in the harder problem difficulty perform poorly (i.e., no consensus or wrong consensus is reached). 
This indicates that evolving decision-making mechanisms in easier problem difficulties is advantageous and allows evolution to come up with more efficient decision-making mechanisms. 

Our hybrid fitness function~$F_{HB}$ led to competitive, but not more efficient, decision-making mechanisms. 
In contrast to our hypothesis, the reward for prediction accuracy did not lead to higher decision speed (i.e., shorter mean consensus time~$\overline{T}_N$). 
The inclusion of the prediction task seems to complicate the setting in such a way that the performance is partly even worse than in the scenario with task-specific fitness function~$F_{TS}$. 
Differing from the results for the task-specific fitness function, the harder problem setting leads to improvements in mean consensus time~$\overline{T}_N$ and exit probability~$E_N$ over the easier problem setting when using the hybrid fitness function.
But the ranges of mean consensus time~$\overline{T}_N$ and exit probability~$E_N$ are larger for the best-evolved individuals in problem difficulty~$\rho^* = 0.52$ than for the best-evolved individuals in problem difficulty~$\rho^* = 0.25$ indicating a larger performance variability. 
Two best-evolved individuals do not lead to a consensus in problem difficulties $\rho^* \in \{0.52,0.67,0.82\}$ and one best-evolved individual does not scale well with problem difficulty (i.e., it does not reach a consensus in harder problem difficulties). 
Still, the overall increase in decision efficiency indicates that including our task-independent reward for prediction accuracy can improve efficiency when the difficulty of the overall setting is high. 

\section{Conclusion} \label{sec:conclusion}

Our experiments show that collective decision-making mechanisms can be evolved in the collective perception scenario using methods of evolutionary computation and task-specific fitness functions or hybrid fitness functions combining a task-specific and a task-independent reward. 
Our task-specific reward is simple and rewards the robot swarm's opinion in the last time step of an evaluation. 
Still, this reward is sufficient to evolve competitive collective decision-making mechanisms that are efficient being accurate and fast. 
Similarly, the best-evolved individuals of the evolutionary runs using our hybrid fitness function are efficient collective decision-making mechanisms. 
All evolved collective decision-making mechanisms outperform the voter model and are slightly slower but more accurate than the majority rule. 
But using a task-independent reward for prediction accuracy did not lead to the emergence of collective decision-making mechanisms. 
Instead, robot behaviors with fixed robot opinions independent from the actual best option emerge allowing trivial predictions of the robot sensors. 

To the best of our knowledge, we are the first to study the evolution of collective decision-making mechanisms using task-independent rewards. 
Although using only a task-independent reward did not lead to the emergence of collective decision-making mechanisms and combining it with a task-specific reward did not lead to more efficient decision-making mechanisms, we still see great potential in evolving collective decision-making mechanisms using task-independent rewards. 
Our experiments with the hybrid fitness function have shown that our task-independent reward for prediction accuracy can be beneficial in complex task settings (see Tab.~\ref{tab:metrics}), as created in our experiments when combining the prediction task with the collective perception task. 
In future work, we will test different patterns of the two environmental features~\cite{bartashevich2019} and switch from the best-of-two setting used here 
to a best-of-n setting with $n > 2$ to increase the difficulty of the collective perception task. 
In these settings, we will investigate if our hybrid fitness function will lead to the emergence of better-performing collective decision-making mechanisms than the task-specific fitness function. 
Another option to increase the overall task difficulty is to evolve the complete decision-making behavior (i.e., we would replace the PFSM for decision-making, see Fig.~\ref{fig:psfm_decision}, with a neural network) or even both decision-making behavior and robot motion (i.e., replacing both finite state machines, see Fig.~\ref{fig:fsms}).  

We will also intensify our studies on using a purely task-independent reward for evolving collective decision-making behaviors. 
The choice of the predicted sensor values potentially has a significant impact on the resulting behaviors. 
Currently, the sensor values giving a robot's neighbor opinions and the decision-making are independent from the sensor values giving input about the quality of the options in the environment.  
A careful configuration of the predictor ANN (see Fig.~\ref{fig:predictor}) and the robot's sensors may enable the emergence of decision-making mechanisms in our purely task-independent setup. 
Another promising approach is to use Markov blankets for calibrating the robots' sampling and communication frequencies to adjust the difficulty of the scenario so that decision-making mechanisms may emerge. 

In addition to studying the influence of the task-independent rewards on the evolved decision-making mechanisms in more detail in future work, we will also study the evolved decision-making strategies in depth to compare their approach to decision-making to the voter model and the majority rule.  


\end{document}